\documentclass[aps, prb, amsmath, twocolumn, longbibliography, 10pt]{revtex4-1}

\usepackage{graphicx} % To import figures
\usepackage{multibib} % Required for multiple bibliographies. Remove if there is a single bibliography
\usepackage{xcolor}
\newcites{supp}{ } % Defines a new \cite command to refer to papers in supp info.
\begin{document}

\title{Dark current in monolithic extended-SWIR GeSn PIN photodetectors}

\author{Mahmoud R. M. Atalla}
\affiliation{Department of Engineering Physics, \'Ecole Polytechnique de Montr\'eal, C.P. 6079, Succ. Centre-Ville, Montr\'eal, Qu\'ebec, Canada H3C 3A7}
\author{Simone Assali}
\affiliation{Department of Engineering Physics, \'Ecole Polytechnique de Montr\'eal, C.P. 6079, Succ. Centre-Ville, Montr\'eal, Qu\'ebec, Canada H3C 3A7}
\author{Sebastian Koelling}
\affiliation{Department of Engineering Physics, \'Ecole Polytechnique de Montr\'eal, C.P. 6079, Succ. Centre-Ville, Montr\'eal, Qu\'ebec, Canada H3C 3A7}
\author{Anis Attiaoui}
\affiliation{Department of Engineering Physics, \'Ecole Polytechnique de Montr\'eal, C.P. 6079, Succ. Centre-Ville, Montr\'eal, Qu\'ebec, Canada H3C 3A7}
\author{Oussama Moutanabbir}
\email{oussama.moutanabbir@polymtl.ca}
\affiliation{Department of Engineering Physics, \'Ecole Polytechnique de Montr\'eal, C.P. 6079, Succ. Centre-Ville, Montr\'eal, Qu\'ebec, Canada H3C 3A7}

\begin{abstract}
The monolithic integration of extended short-wave infrared (e-SWIR) photodetectors (PDs) on silicon is highly sought-after to implement manufacturable, cost-effective sensing and imaging technologies. With this perspective, GeSn PIN PDs have been the subject of extensive investigations because of their bandgap tunability and silicon compatibility. However, due to growth defects, these PDs suffer a relatively high dark current density as compared to commercial III-V PDs. Herein, we elucidate the mechanisms governing the dark current in $2.6 \, \mu$m GeSn PDs at a Sn content of $10$ at.\%. It was found that in the temperature range of $293 \, $K -- $363 \,$K and at low bias, the diffusion and Shockley-Read-Hall (SRH) leakage mechanisms dominate the dark current in small diameter ($20 \, \mu$m) devices, while combined SRH and trap assisted tunneling (TAT) leakage mechanisms are prominent in larger diameter ($160 \, \mu$m) devices. However, at high reverse bias, TAT leakage mechanism becomes dominant regardless of the operating temperature and device size. The effective non-radiative carrier lifetime in these devices was found to reach $\sim 300$ -- $400$ ps at low bias. Owing to TAT leakage current, however, this lifetime reduces progressively as the bias increases.
\end{abstract}

\maketitle

GeSn semiconductors have been attracting a great deal of interest because of the flexibility they offer to engineer the lattice parameter and the bandgap energy and directness \cite{moutanabbir2021monolithic,elbaz2020ultra,zhou2020electrically,ch2019gesn}. In recent years, silicon-integrated GeSn photodetectors (PDs) were demonstrated showing a room temperature performance close to that of commercial PbSe detectors at wavelengths reaching $3 \,\mu$m \cite{kim2022enhanced,lin2021temperature}. However, GeSn semiconductors are inherently metastable and typically exhibit a compositional gradient and a large compressive strain resulting from the lattice- mismatched growth on Ge/Si substrates \cite{werner2011germanium,al2016study,assali2019enhanced}. As a matter of fact, GeSn PDs suffer high dark leakage current, which can be orders of magnitude higher than that of extended-InGaAs PDs. For instance, at a wavelength cutoff of $2.5 \,\mu$m GeSn PIN diodes display dark current density of $\sim 10$ A/cm$^2$, whereas in commercial InGaAs PIN PDs this density is typically on the order of $1$ mA/cm$^2$ \cite{xu2019high,tran2019si,atalla2021high}. This reduces the performance of GeSn PDs by limiting the detectivity and the noise equivalent power, thus hindering their use in optical communication applications where low dark current PDs are required \cite{atalla2021all,yang2019highly}. High performance, cost-effective e-SWIR detectors operating at room temperature at wavelengths above 2 µm are also coveted for time-resolved spectroscopy, surveillance, autonomous vehicles, and imaging through scattering media such as snow, haze, and fog \cite{atalla2021high,soref2015enabling,wang2021high}. 

The lattice mismatch between Ge and GeSn layers creates misfits and threading dislocations, which are the main reason for the high dark current \cite{simoen2007germanium,son2020dark,dong2017two,zhou2020high}. However, studies of the dominating leakage mechanisms and the non-radiative carrier lifetime in GeSn PIN photodetectors remain conspicuously missing in literature despite their crucial importance to the development of all-group IV e-SWIR technologies. In this vein, herein dark current analysis is performed for epitaxial PIN GeSn PD devices at $10$ at.\% Sn content in the i-layer corresponding to a wavelength cutoff of 2.6 µm.  For device diameters between $20 \, \mu$m to $160 \, \mu$m, it was found that bulk leakage rather than surface leakage is dominating the dark current, especially as the device diameter increases. Temperature-dependent dark current measurements were analyzed to estimate the activation energies as a function of the reverse bias. It was found that, for the small diameter device, diffusion dominates at low reverse bias, whereas trap assisted tunnel (TAT) leakage mechanism dominates at high bias. The latter is also the case for the large diameter device, but at low reverse bias Shockley-Read-Hall (SRH) leakage becomes the prevalent mechanism. Using capacitance-voltage measurements, the effective non-radiative carrier lifetime estimated at low bias was found in the 300 - 400 ps range, which decreases as high bias due TAT leakage dominating the dark current.

\bigskip

%\section{Results and discussion}

% Fig.~\ref{fig:fig1}%
\begin{figure}
\includegraphics[width=8.2cm, height=6.0cm]{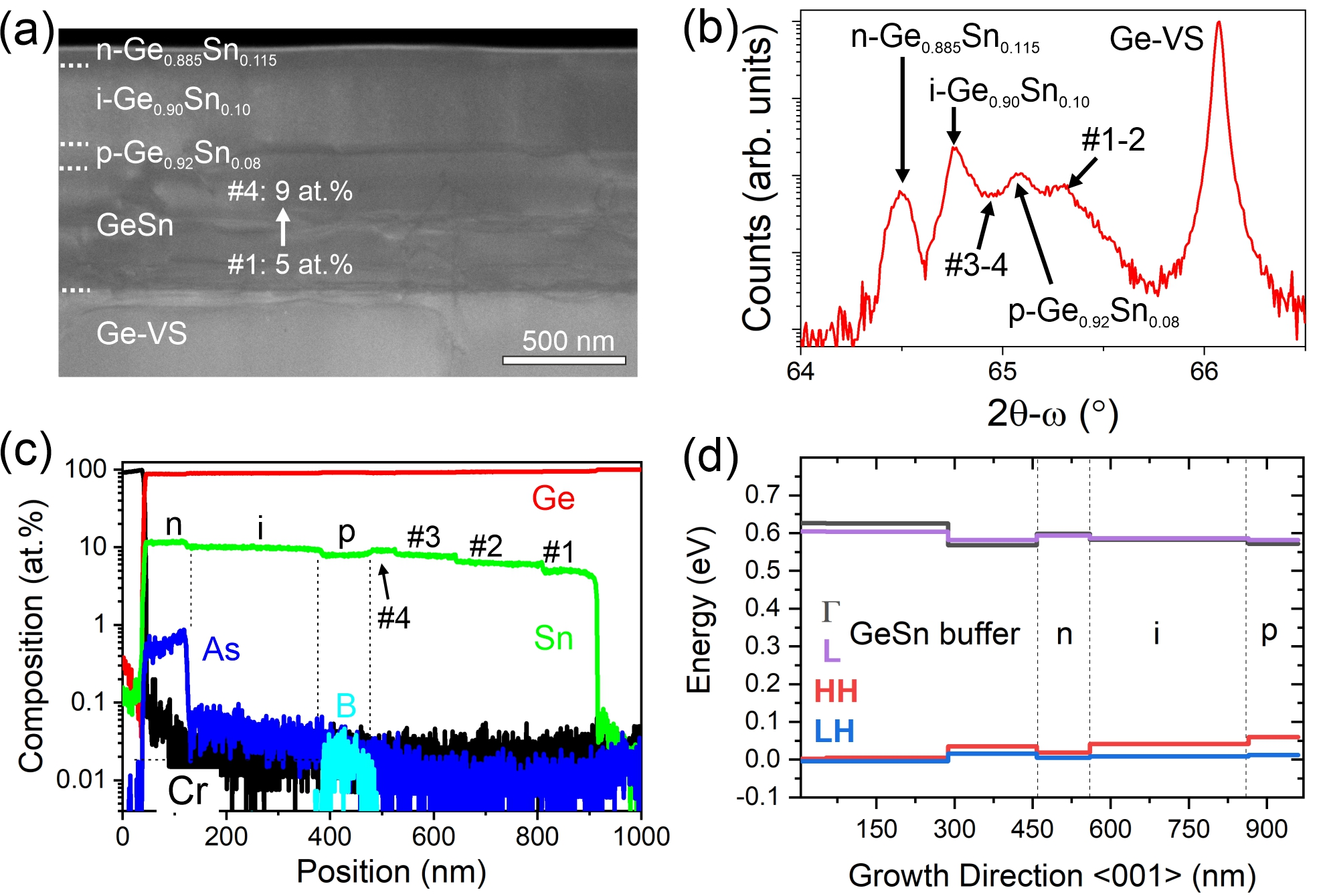}% Here is how to import EPS art
\caption{\label{fig:fig1}Material growth and characterization. (a) TEM image showing the CVD grown PIN stack on top of GeSn and Ge-VS buffer layers. (b) (XRD) $2 \theta - \omega$ scan confirming the high crystalline quality of GeSn PIN stack. (c) APT concentration profiles of different elements across the GeSn stack showing the width of the i-layer to be $300\,\text{nm}$. (d) Calculated $8 \times 8 \, \, k \cdot p$ band lineup (at $300$ K) for the whole Ge VS/GeSn stack including the PIN layers.}
\end{figure}

In a low-pressure chemical vapor deposition (CVD) reactor, a $1.5\,\mu$m Ge virtual substrate (VS) was grown on a 4-inch Si ($100$) wafer. Afterward, GeSn multi-layer buffer with step-wise Sn content increase was grown on Ge VS to engineer the lattice constant to allow for the growth of PIN GeSn heterostructure at the desired Sn composition and lattice strain. A cross-sectional transmission electron micrograph (TEM) is displayed in Fig. 1(a) showing the heterostructure indicating a total thickness of $960$ nm. Threading dislocations are observed in the Ge VS/GeSn buffer layers and p-GeSn, however, no extended defects are observed in the i-GeSn and n-GeSn, at the TEM imaging scale. This is likely because of the nucleation and propagation of threading dislocations are suppressed and the gliding of misfit dislocation is promoted instead at the interface of GeSn buffer layers underneath the PIN heterostructure \cite{assali2019enhanced}.

X-ray diffraction spectroscopy (XRD) $2\theta - \omega$ scans around the ($004$) XRD order were performed to further investigate the crystalline quality of as-grown GeSn multi-layer stack, as shown in Fig. 1(b). The Ge-VS peak is detected at $66.06 ^{\circ}$, while the signal at smaller angles relates to the GeSn stack. The n-GeSn/i-GeSn/p-GeSn peaks are observed at $64.51 ^{\circ}$, $64.76 ^{\circ}$ and  $65.08 ^{\circ}$, respectively. Since there are four GeSn buffer layers, the layers \# 1 -- 2 have a peak around $65.31 ^{\circ}$ and the layers \# 3 -- 4 are observed around $64.94 ^{\circ}$ \cite{atalla2021high}. The reported peak at $65.8 ^{\circ}$ associated with severe Sn segregation and precipitation \cite{aubin2017growth,aubin2017impact} is not observed thus confirming the absence of Sn clusters, in agreement with the atom probe tomography data (APT) measurements discussed below. The high intensity narrow n-GeSn/i-GeSn peaks indicates the high crystalline quality of these layers in line with TEM analysis. To determine the Sn content and investigate the p- and n-doping concentrations, three-dimensional atom-by-atom APT mapping was performed, as shown in Fig. 1(c). Although the small difference in atomic mass between As and Ge atoms makes decoupling As from Ge signal in APT challenging, a clear As profile is visible in the top GeSn layer with a background signal across the heterostructure. The Sn incorporation is controlled by a  $10 ^{\circ}$C-step temperature reduction during growth resulting in uniform compositions of $5$ at.\% (\#1), $6$ at.\% (\#2), $8$ at.\% (\#3), and $9$ at.\% (\#4). Owing to the relatively slow B incorporation in GeSn lattice, the Sn content in the p-GeSn layer decreases from 9 to 8 at.\% with broader interface. Conversely, the Sn content increases by $1$ at.\% in n-Ge$_{0.885}$Sn$_{0.115}$ (Fig. 1(c)), which stands in sharp contrast to earlier observations, where As doping was reported to reduce the incorporation of Sn. \cite{senaratne2014advances,bhargava2017doping,margetis2017fundamentals}

\bigskip

Since both strain and composition influence the GeSn bandstructure, the eight-band $k\cdot p$ model with the envelope function approximation was optimized using the empirical parameters extracted from Fig. 1(a-c) to calculate the band lineup of the whole stack \cite{assali2021midinfrared}, as depicted in Fig.\ref{fig:fig1}(d). It is noteworthy that the n-Ge$_{0.885}$Sn$_{0.115}$ and i-Ge$_{0.894}$Sn$_{0.106}$ layers have direct bandgaps of $0.512$ eV and $0.541$ eV, respectively. The p-Ge$_{0.924}$Sn$_{0.076}$ layer displays an indirect bandgap of $0.575$ eV. It is also noted that the tensile strain in the n-layer has reduced the bandgap energy as compared to the i-layer which possesses less Sn content.

\bigskip

% Fig.~\ref{fig:fig2}%
\begin{figure}
\includegraphics[width=8cm, height=6cm]{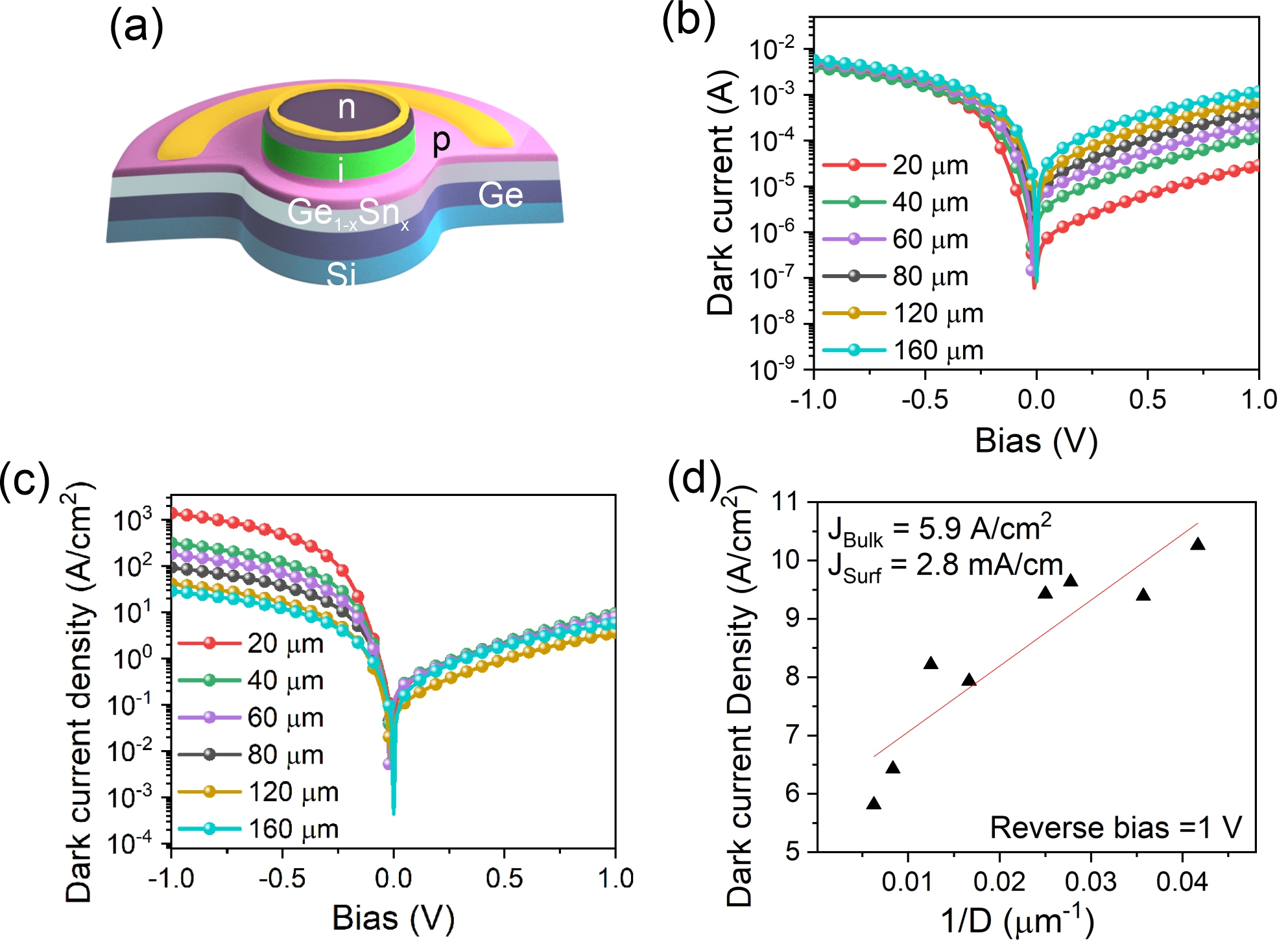}% Here is how to import EPS art
\caption{\label{fig:fig2}GeSn PIN photodetectors (PDs). (a) Schematic illustration of the microfabricated PDs. (b) The IV characteristics in  dark for various device diameters. (c) Dark current density of GeSn PIN PDs. (d) Dark current density as a function of the inverse of the device diameter at a reverse bias of $1$ V. The surface and bulk current densities can be extracted from the linear fit.}
\end{figure}

% Fig.~\ref{fig:fig3}%
\begin{figure*}
\centering
\includegraphics[width=14cm, height=7cm]{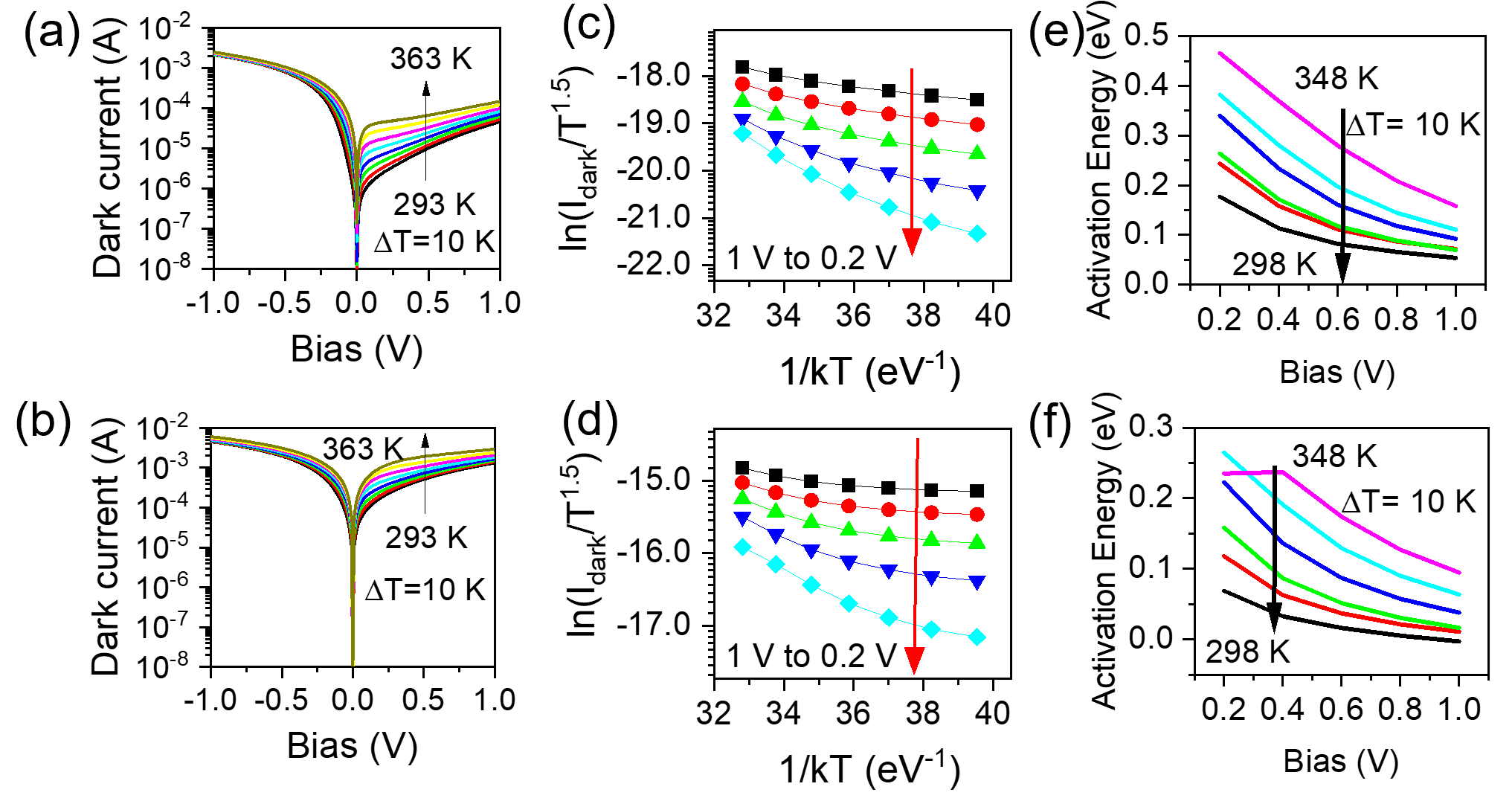}% Here is how to import EPS art
\caption{\label{fig:fig3}(a) and (b) temperature dependent IV measurements for $20 \,\mu$m and $160 \,\mu$m devices, respectively. (c) and (d) The Arrhenius plots for $20 \,\mu$m and $160 \,\mu$m devices at variable bias up to 1 V with a 0.2 V step, respectively. (e) and (f) The obtained activation energy of GeSn PDs estimated from temperature-dependent dark current measurements}
\end{figure*}

% Fig.~\ref{fig:wide}%

 PIN PDs were fabricated with various mesa diameters following a top-down processing flow. Fig. 2(a) illustrates the device design. To reduce the contact parasitic capacitance and isolate each device from its neighboring ones, the fabrication process started with chlorine ICP etch down to Ge-VS. Subsequently, Another chlorine etch in the shape of circular bumps was subsequently performed down to the p-layer. The fabricated devices have varying mesa diameters of $20$, $24$, $32$, $40$, $80$, $120$, and $160$ $\mu$m. To reduce surface defects, the etched sidewall was passivated using chemical treatment in HCl : HF (1:1) for 10 seconds followed by a PECVD deposited thick SiO$_{2}$ layer. Using BOE wet etch, openings were made in the SiO$_{2}$ layer for the p- and n-contacts followed by Ti/Au contacts deposition using e-beam evaporation, as schematically shown in Fig. \ref{fig:fig2}(a). 

First, the IV curves of GeSn PIN PDs at different diameters in dark were investigated. As displayed in Fig. 2b, a rectification ratio of $\> 10^2$ is recorded at $0.5$ V for the 20 µm device, which indicates the relatively high quality of the GeSn PIN layers. As the diameter size increases, the reverse dark current increases monotonically, which is expected because of the increased active area of the device. However, when the dark current density is plotted as a function of the applied bias, the reverse dark current densities overlap on each other regardless of the device diameter, which is indicative of the low surface current leakage as the active area reduces from $160 \,\mu$m to just $20 \,\mu$m, as shown in Fig. \ref{fig:fig2}(c). It is worth mentioning that the low dark current and applied bias would significantly increase the detectivity of these devices compared to photoconductive (PC) devices that suffer from high dark current and high noise. This would allow PIN devices to work without the need for lock-in technique to extract the photocurrent signal, which is relevant for practical applications.
Owing to the importance of the dark current in shaping the performance of a PD, in-depth investigations of the underlying processes are of compelling importance. In this vein, systematic studies of the temperature-dependent dark current, activation energy, and leakage mechanisms are performed. One way to separate the contributions of the bulk leakage from the surface leakage is to fit the dark current density $J_{dark}$ as a function of the inverse of the device diameter ($D$) using \cite{zhou2020high}: 

\begin{equation}\label{Jdeq}
    J_{dark}= J_{bulk}+4\,J_{surf}/\text{D},
\end{equation}

\noindent where $J_{surf}$ is the surface leakage current density. Figure \ref{fig:fig2}(d) shows the linear fit for $J_{dark}$ at reverse bias of $1$ V and the value of the  $J_{bulk}$ is $6.5$ A$/$cm$^2$ which is $3$ orders of magnitude higher than the $J_{surf}$ of $2$ mA$/$cm$^2$ indicating the effective passivation of these PIN PDs.  
Thus, only the bulk leakage current is considered in the following analysis as it is the dominating component. 
% TEM section and defects discussion as in Fig 5b 

%\textcolor{red}{search for temp. dependent CV meas.}

In Figs. \ref{fig:fig3}(a) and \ref{fig:fig3}(b), the dark current is plotted for the smallest and largest devices as a function of the applied bias acquired at different temperatures in the $293$ - $363$ K range with $10$ K increments. At low bias ($< \,0.5$ V), it is noticeable that the rise in temperature yields a remarkable increase in the dark current in the small device as compared to the large one. However, at high bias ($> \, 0.5$ V) the dark current at various temperatures becomes practically identical regardless of the device size. The following analysis elucidates this behavior.

Based on the temperature-dependent studies, the dominant leakage mechanisms underlying the dark current are identified, and their activation energies are estimated. According to PL measurements (not shown) the effective bandgap for the GeSn device is $0.52$ eV ($2.4\,\mu$m) \cite{atalla2021high}. Generally, the dark current generation in PIN device is attributed to three main mechanisms \cite{dilello2012characterization,son2020dark}: (1) diffusion leakage current due to the minority carriers at the edges of the depletion region, which corresponds to an activation energy (E$_a$) close the bandgap energy ($\sim 0.52$ eV); (2) SRH leakage current caused by the generation of carriers from the deep level defect states with the corresponding activation energy of about half of the bandgap energy $\sim 0.26$ eV; and (3) TAT leakage current, which is more pronounced under a significantly high reverse bias, and the associated activation energy is typically smaller than that of SRH (< $0.26$ eV). Given the relatively low reverse bias applied to these GeSn PDs and the very high bias typically required for the band-to-band tunneling, the contribution of the latter is neglected under our conditions. Consequently, the dark current can be expressed as:

\begin{equation}\label{Ideq}
    I_{dark} = I_{diff} + I_{SRH} + I_{TAT},
\end{equation}

\noindent where $I_{diff}$, $I_{SRH}$, $I_{TAT}$ are the diffusion, SRH, and TAT dark current components, respectively. Since the $I_{TAT}$ depends on the trap density of the bulk material, it can be expressed as $I_{TAT} = \Gamma \, I_{SRH}$, where $\Gamma$ is the simulated electric field enhancement factor, which is proportional to the average electric field inside the junction. \cite{hurkx1992new,gonzalez2011analysis}

\bigskip

The total leakage dark current can be expressed as: \cite{sze2021physics}

\begin{equation}\label{Ideq}
    I_{dark}=BT^{1.5} e^{-E_a/kT} (e^{qV/2kT}-1),
\end{equation}

\noindent where $B$ is a constant and $V$ is the applied bias. The Arrhenius plot of ln$\,$(I$_{dark}$/T$^{1.5}$) as function of $1/kT$ for various bias values of $1$, $0.8$, $0.6$, $0.4$, and $0.2$ V is shown in Figs. \ref{fig:fig3}(c) and \ref{fig:fig3}(d). It can be inferred from this plot that the curves become more linear as the bias increases for the small size device, which indicates a change in the leakage current mechanism, as discussed below.

For $qV >> kT$, the activation energy is the slope of these curves. The obtained values are plotted as function of the applied bias for the two set of devices in Figs. \ref{fig:fig3}(e) and \ref{fig:fig3}(f). Two different regimes emerge from this analysis. First, at low bias $< 0.5$ V, for the small diameter device E$_a$ significantly increases with temperature to reach $\sim \,0.5$ eV at $0.2$ V. This indicates that the diffusion leakage current is dominating at high T, while the TAT current dominates at low temperature, where E$_a < \text{E}_g/$2. However, for the large diameter device E$_a \sim \, 0.26$ eV (i.e, $\text{E}_g/$2) is obtained at high temperature indicating that the SRH generation mechanism is dominating the leakage current. At low temperature, the TAT  dominates the dark current. This is most likely because the large active area devices contain a higher absolute amount of dislocations and defects as compared to the small ones. Second, at high bias $> 0.5$ V, the E$_a << \text{E}_g/$2 indicate that TAT is dominating the leakage current regardless of the device size or temperature. This is due to the relatively thin i-layer of these GeSn PDs such as only $1$ V is sufficient to fully deplete it and enforce a high tunneling current.

\begin{figure}
\includegraphics[width=8.5cm, height=3.2cm]{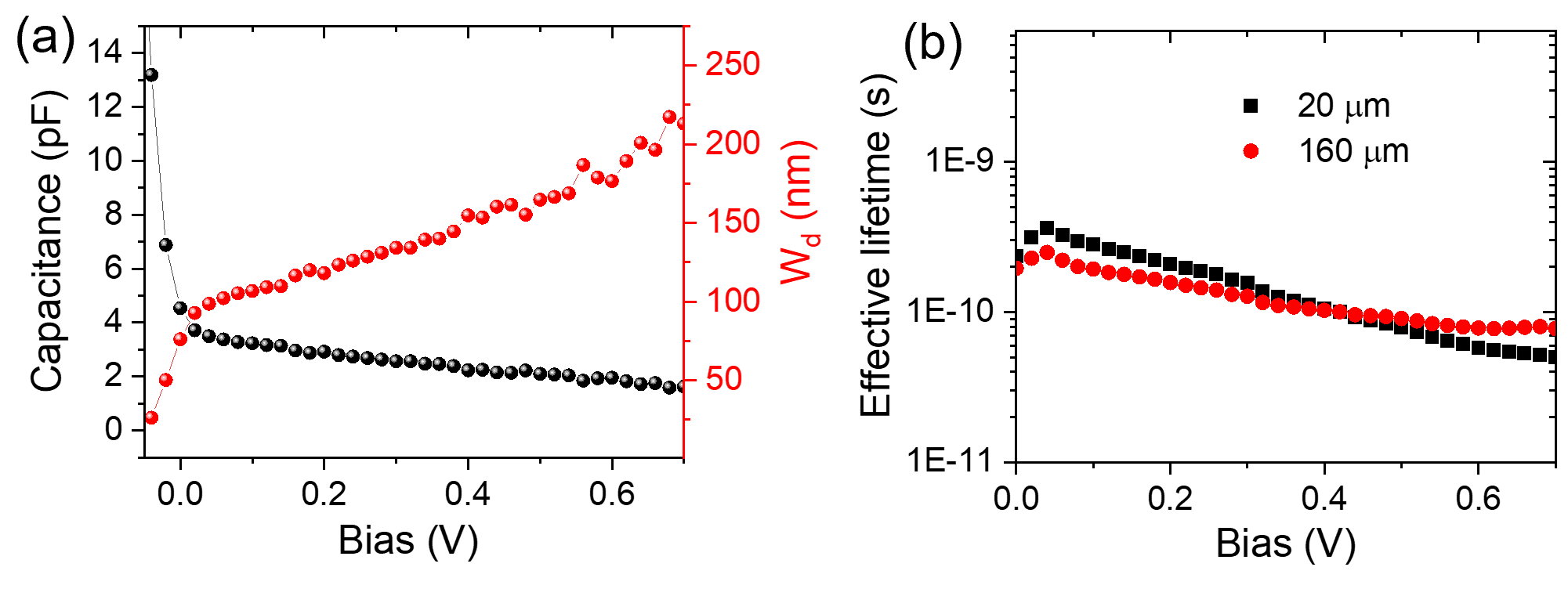}% Here is how to import EPS art
\caption{\label{fig:fig4}(a) Measured capacitance and calculated depletion width $W$ as a function of bias for the $20 \, \mu$m diameter device, and (b) the estimated carrier lifetime as function of bias for the $20 \mu$m and $160 \, \mu$m devices.}
\end{figure}

To further clarify the leakage processes in these PIN devices, it is important to estimate the effective carrier lifetime $\tau _{eff}$. The latter is obtained assuming that the diffusion length is longer than the depletion region \cite{vsvcajev2020temperature} and by differentiating the dark current by the depletion width ($W$) as

\begin{equation}\label{effeq}
    \frac{dJ_{dark}}{dW} \approx \frac{q\,n_i}{ \tau _{eff}},
\end{equation}

\noindent where n$_i$ is the intrinsic carrier concentration in the i-layer. It is relevant to evaluate the dependence of $\tau_{eff}$ on the bias and device size when TAT is not the dominant mechanism. Based on MOS capacitor measurements, the estimated carrier concentration in the unintentionally doped as-grown GeSn thin films is n$_i = 1.05\,\times10^{17}$ cm$^{-3}$ at room temperature. To determine the $W$, capacitance-voltage measurements (CV) were performed for all devices and the depletion width was determined as $W = \, \epsilon A/C$, where $A$ is the active device area and $\epsilon$ is the dielectric permittivity of GeSn. The depletion width is assumed to be independent of the device diameter. As shown in Fig. \ref{fig:fig4}(a), the capacitance rapidly drops close to $0$ V then slowly decreases as the reverse bias increases. The $W$ extracted from the capacitance curve shows a continuous increase and exceeds $200$ nm at $0.7$ V beyond which the capacitance values are depressed because of the significant TAT leakage. Knowing $W$, $\tau_{eff}$ is determined from Eq.\ref{effeq} as displayed in Fig. \ref{fig:fig4}(b) showing $\tau_{eff}$ as a function of the applied bias up to $0.7$ V for both small and large devices at T = 293 K. It can be inferred from the figure that, regardless of the device size, $\tau_{eff}$ decreases as the bias increases. Note that the activation energies in Figs. \ref{fig:fig3}(e) and \ref{fig:fig3}(f) are $<<E_g / 2$. This indicates that TAT leakage dominates the dark current at room temperature, which explains the small values of $\tau_{eff}$ across the whole bias range. At a low bias $< 0.5$ V, owing to the increased bulk defects in the large device and the increased importance of surface defects in the small device, and that $\Gamma$depends on the electric field, the small device has slightly larger $\tau_{eff}$. Conversely, at a high bias $> 0.5$ V the large device exhibits a slightly higher $\tau_{eff}$. This is because the electric field has greater influence on the small device surface as the bias increases. However, this surface effect remains rather minor.

In summary, the dark current characteristics of vertical PIN GeSn PDs was investigated. The various leakage mechanisms dominating the dark current were elucidated and their behavior as a function of bias, device dimension, and operating temperature were discussed. Additionally, the effective non-radiative carrier lifetime was found to exceed $300$ ps at low bias and decreases monotonically as the bias increases due to TAT leakage. Understanding and controlling the dark current is of paramount importance to pave the way to improve GeSn e-SWIR PDs and adopte their use in optical and data communication, LIDAR, and biochemical sensing applications.

\bigskip

\noindent {\bf Acknowledgements}.
The authors thank J. Bouchard for the technical support with the CVD system, O.M. acknowledges support from NSERC Canada (Discovery, SPG, and CRD Grants), Canada Research Chairs, Canada Foundation for Innovation, Mitacs, PRIMA Québec, and Defence Canada (Innovation for Defence Excellence and Security, IDEaS).

\medskip

\medskip
\noindent {\bf Author information}.
Correspondence and requests for materials should be addressed to~:\\ oussama.moutanabbir@polymtl.ca

\medskip
\noindent {\bf Data availability}.
The data that support the findings of this study are available from the corresponding author upon reasonable request.

\bibliographystyle{naturemag}
\bibliography{main}

\begin{thebibliography}{10}
\expandafter\ifx\csname url\endcsname\relax
  \def\url#1{\texttt{#1}}\fi
\expandafter\ifx\csname urlprefix\endcsname\relax\def\urlprefix{URL }\fi
\providecommand{\bibinfo}[2]{#2}
\providecommand{\eprint}[2][]{\url{#2}}

\bibitem{moutanabbir2021monolithic}
\bibinfo{author}{Moutanabbir, O.} \emph{et~al.}
\newblock \bibinfo{title}{Monolithic infrared silicon photonics: the rise of
  \text{(Si)GeSn} semiconductors}.
\newblock \emph{\bibinfo{journal}{Applied Physics Letters}}
  \textbf{\bibinfo{volume}{118}}, \bibinfo{pages}{110502}
  (\bibinfo{year}{2021}).

\bibitem{elbaz2020ultra}
\bibinfo{author}{Elbaz, A.} \emph{et~al.}
\newblock \bibinfo{title}{Ultra-low-threshold continuous-wave and pulsed lasing
  in tensile-strained \text{GeSn} alloys}.
\newblock \emph{\bibinfo{journal}{Nature Photonics}}
  \textbf{\bibinfo{volume}{14}}, \bibinfo{pages}{375--382}
  (\bibinfo{year}{2020}).

\bibitem{zhou2020electrically}
\bibinfo{author}{Zhou, Y.} \emph{et~al.}
\newblock \bibinfo{title}{Electrically injected \text{GeSn} lasers on \text{Si}
  operating up to 100 \text{K}}.
\newblock \emph{\bibinfo{journal}{Optica}} \textbf{\bibinfo{volume}{7}},
  \bibinfo{pages}{924--928} (\bibinfo{year}{2020}).

\bibitem{ch2019gesn}
\bibinfo{author}{Chretien, J.} \emph{et~al.}
\newblock \bibinfo{title}{\text{GeSn} lasers covering a wide wavelength range
  thanks to uniaxial tensile strain}.
\newblock \emph{\bibinfo{journal}{ACS Photonics}} \textbf{\bibinfo{volume}{6}},
  \bibinfo{pages}{2462--2469} (\bibinfo{year}{2019}).

\bibitem{kim2022enhanced}
\bibinfo{author}{Kim, Y.} \emph{et~al.}
\newblock \bibinfo{title}{Enhanced \text{GeSn} microdisk lasers directly
  released on \text{Si}}.
\newblock \emph{\bibinfo{journal}{Advanced Optical Materials}}
  \textbf{\bibinfo{volume}{10}}, \bibinfo{pages}{2101213}
  (\bibinfo{year}{2022}).

\bibitem{lin2021temperature}
\bibinfo{author}{Lin, K.-C.}, \bibinfo{author}{Huang, P.-R.},
  \bibinfo{author}{Li, H.}, \bibinfo{author}{Cheng, H.} \&
  \bibinfo{author}{Chang, G.-E.}
\newblock \bibinfo{title}{Temperature-dependent characteristics of
  \text{GeSn/Ge} multiple-quantum-well photoconductors on silicon}.
\newblock \emph{\bibinfo{journal}{Optics Letters}}
  \textbf{\bibinfo{volume}{46}}, \bibinfo{pages}{3604--3607}
  (\bibinfo{year}{2021}).

\bibitem{werner2011germanium}
\bibinfo{author}{Werner, J.} \emph{et~al.}
\newblock \bibinfo{title}{Germanium-tin pin photodetectors integrated on
  silicon grown by molecular beam epitaxy}.
\newblock \emph{\bibinfo{journal}{Applied Physics Letters}}
  \textbf{\bibinfo{volume}{98}}, \bibinfo{pages}{061108}
  (\bibinfo{year}{2011}).

\bibitem{al2016study}
\bibinfo{author}{Al-Kabi, S.} \emph{et~al.}
\newblock \bibinfo{title}{Study of high-quality \text{GeSn} alloys grown by
  chemical vapor deposition towards mid-infrared applications}.
\newblock \emph{\bibinfo{journal}{Journal of Electronic Materials}}
  \textbf{\bibinfo{volume}{45}}, \bibinfo{pages}{6251--6257}
  (\bibinfo{year}{2016}).

\bibitem{assali2019enhanced}
\bibinfo{author}{Assali, S.}, \bibinfo{author}{Nicolas, J.} \&
  \bibinfo{author}{Moutanabbir, O.}
\newblock \bibinfo{title}{Enhanced \text{Sn} incorporation in \text{GeSn}
  epitaxial semiconductors via strain relaxation}.
\newblock \emph{\bibinfo{journal}{Journal of Applied Physics}}
  \textbf{\bibinfo{volume}{125}}, \bibinfo{pages}{025304}
  (\bibinfo{year}{2019}).

\bibitem{xu2019high}
\bibinfo{author}{Xu, S.} \emph{et~al.}
\newblock \bibinfo{title}{High-speed photodetection at two-micron-wavelength:
  technology enablement by \text{GeSn/Ge} multiple-quantum-well photodiode on
  300 mm \text{Si} substrate}.
\newblock \emph{\bibinfo{journal}{Optics express}}
  \textbf{\bibinfo{volume}{27}}, \bibinfo{pages}{5798--5813}
  (\bibinfo{year}{2019}).

\bibitem{tran2019si}
\bibinfo{author}{Tran, H.} \emph{et~al.}
\newblock \bibinfo{title}{Si-based \text{GeSn} photodetectors toward
  mid-infrared imaging applications}.
\newblock \emph{\bibinfo{journal}{ACS Photonics}} \textbf{\bibinfo{volume}{6}},
  \bibinfo{pages}{2807--2815} (\bibinfo{year}{2019}).

\bibitem{atalla2021high}
\bibinfo{author}{Atalla, M.}, \bibinfo{author}{Assali, S.},
  \bibinfo{author}{Koelling, S.}, \bibinfo{author}{Attiaoui, A.} \&
  \bibinfo{author}{Moutanabbir, O.}
\newblock \bibinfo{title}{High-bandwidth extended-\text{SWIR} \text{GeSn}
  photodetectors on silicon achieving ultrafast broadband spectroscopic
  response}.
\newblock \emph{\bibinfo{journal}{arXiv preprint arXiv:2111.02892}}
  (\bibinfo{year}{2021}).

\bibitem{atalla2021all}
\bibinfo{author}{Atalla, M.~R.} \emph{et~al.}
\newblock \bibinfo{title}{All-group \text{IV} transferable membrane
  mid-infrared photodetectors}.
\newblock \emph{\bibinfo{journal}{Advanced Functional Materials}}
  \textbf{\bibinfo{volume}{31}}, \bibinfo{pages}{2006329}
  (\bibinfo{year}{2021}).

\bibitem{yang2019highly}
\bibinfo{author}{Yang, F.} \emph{et~al.}
\newblock \bibinfo{title}{Highly enhanced \text{SWIR} image sensors based on
  \text{Ge$_{1-x}$Sn$_x$} graphene heterostructure photodetector}.
\newblock \emph{\bibinfo{journal}{ACS Photonics}} \textbf{\bibinfo{volume}{6}},
  \bibinfo{pages}{1199--1206} (\bibinfo{year}{2019}).

\bibitem{soref2015enabling}
\bibinfo{author}{Soref, R.}
\newblock \bibinfo{title}{Enabling 2 $\mu$m communications}.
\newblock \emph{\bibinfo{journal}{Nature Photonics}}
  \textbf{\bibinfo{volume}{9}}, \bibinfo{pages}{358--359}
  (\bibinfo{year}{2015}).

\bibitem{wang2021high}
\bibinfo{author}{Wang, H.} \emph{et~al.}
\newblock \bibinfo{title}{High-speed and high-responsivity pin waveguide
  photodetector at a 2 $\mu$m wavelength with a \text{Ge}$_{0.92}$
  \text{Sn}$_{0.08}$/\text{Ge} multiple-quantum-well active layer}.
\newblock \emph{\bibinfo{journal}{Optics Letters}}
  \textbf{\bibinfo{volume}{46}}, \bibinfo{pages}{2099--2102}
  (\bibinfo{year}{2021}).

\bibitem{simoen2007germanium}
\bibinfo{author}{Simoen, E.~R.} \emph{et~al.}
\newblock \bibinfo{title}{Germanium content dependence of the leakage current
  of recessed \text{SiGe} source/drain junctions}.
\newblock \emph{\bibinfo{journal}{Journal of Materials Science: Materials in
  Electronics}} \textbf{\bibinfo{volume}{18}}, \bibinfo{pages}{787--791}
  (\bibinfo{year}{2007}).

\bibitem{son2020dark}
\bibinfo{author}{Son, B.}, \bibinfo{author}{Lin, Y.}, \bibinfo{author}{Lee,
  K.~H.}, \bibinfo{author}{Chen, Q.} \& \bibinfo{author}{Tan, C.~S.}
\newblock \bibinfo{title}{Dark current analysis of germanium-on-insulator
  vertical pin photodetectors with varying threading dislocation density}.
\newblock \emph{\bibinfo{journal}{Journal of Applied Physics}}
  \textbf{\bibinfo{volume}{127}}, \bibinfo{pages}{203105}
  (\bibinfo{year}{2020}).

\bibitem{dong2017two}
\bibinfo{author}{Dong, Y.} \emph{et~al.}
\newblock \bibinfo{title}{Two-micron-wavelength germanium-tin photodiodes with
  low dark current and gigahertz bandwidth}.
\newblock \emph{\bibinfo{journal}{Optics express}}
  \textbf{\bibinfo{volume}{25}}, \bibinfo{pages}{15818--15827}
  (\bibinfo{year}{2017}).

\bibitem{zhou2020high}
\bibinfo{author}{Zhou, H.} \emph{et~al.}
\newblock \bibinfo{title}{High-efficiency \text{GeSn/Ge} multiple-quantum-well
  photodetectors with photon-trapping microstructures operating at 2 $\mu$m}.
\newblock \emph{\bibinfo{journal}{Optics express}}
  \textbf{\bibinfo{volume}{28}}, \bibinfo{pages}{10280--10293}
  (\bibinfo{year}{2020}).

\bibitem{aubin2017growth}
\bibinfo{author}{Aubin, J.} \emph{et~al.}
\newblock \bibinfo{title}{Growth and structural properties of step-graded, high
  \text{Sn} content \text{GeSn} layers on \text{Ge}}.
\newblock \emph{\bibinfo{journal}{Semiconductor Science and Technology}}
  \textbf{\bibinfo{volume}{32}}, \bibinfo{pages}{094006}
  (\bibinfo{year}{2017}).

\bibitem{aubin2017impact}
\bibinfo{author}{Aubin, J.} \emph{et~al.}
\newblock \bibinfo{title}{Impact of thickness on the structural properties of
  high tin content \text{GeSn} layers}.
\newblock \emph{\bibinfo{journal}{Journal of Crystal Growth}}
  \textbf{\bibinfo{volume}{473}}, \bibinfo{pages}{20--27}
  (\bibinfo{year}{2017}).

\bibitem{senaratne2014advances}
\bibinfo{author}{Senaratne, C.}, \bibinfo{author}{Gallagher, J.},
  \bibinfo{author}{Aoki, T.}, \bibinfo{author}{Kouvetakis, J.} \&
  \bibinfo{author}{Menendez, J.}
\newblock \bibinfo{title}{Advances in light emission from group-\text{IV}
  alloys via lattice engineering and n-type doping based on custom-designed
  chemistries}.
\newblock \emph{\bibinfo{journal}{Chemistry of Materials}}
  \textbf{\bibinfo{volume}{26}}, \bibinfo{pages}{6033--6041}
  (\bibinfo{year}{2014}).

\bibitem{bhargava2017doping}
\bibinfo{author}{Bhargava, N.}, \bibinfo{author}{Margetis, J.} \&
  \bibinfo{author}{Tolle, J.}
\newblock \bibinfo{title}{As doping of \text{SiGeSn} epitaxial semiconductor
  materials on a commercial \text{CVD} reactor}.
\newblock \emph{\bibinfo{journal}{Semiconductor Science and Technology}}
  \textbf{\bibinfo{volume}{32}}, \bibinfo{pages}{094003}
  (\bibinfo{year}{2017}).

\bibitem{margetis2017fundamentals}
\bibinfo{author}{Margetis, J.} \emph{et~al.}
\newblock \bibinfo{title}{Fundamentals of \text{Ge$_{1-x}$Sn$_x$} and
  \text{Si$_y$Ge$_{1-x-y}$Sn$_x$} \text{RPCVD} epitaxy}.
\newblock \emph{\bibinfo{journal}{Materials Science in Semiconductor
  Processing}} \textbf{\bibinfo{volume}{70}}, \bibinfo{pages}{38--43}
  (\bibinfo{year}{2017}).

\bibitem{assali2021midinfrared}
\bibinfo{author}{Assali, S.} \emph{et~al.}
\newblock \bibinfo{title}{Midinfrared emission and absorption in strained and
  relaxed direct-band-gap ge 1- x sn x semiconductors}.
\newblock \emph{\bibinfo{journal}{Physical Review Applied}}
  \textbf{\bibinfo{volume}{15}}, \bibinfo{pages}{024031}
  (\bibinfo{year}{2021}).

\bibitem{dilello2012characterization}
\bibinfo{author}{DiLello, N.}, \bibinfo{author}{Johnstone, D.} \&
  \bibinfo{author}{Hoyt, J.}
\newblock \bibinfo{title}{Characterization of dark current in \text{Ge-on-Si}
  photodiodes}.
\newblock \emph{\bibinfo{journal}{Journal of Applied Physics}}
  \textbf{\bibinfo{volume}{112}}, \bibinfo{pages}{054506}
  (\bibinfo{year}{2012}).

\bibitem{hurkx1992new}
\bibinfo{author}{Hurkx, G.}, \bibinfo{author}{Klaassen, D.} \&
  \bibinfo{author}{Knuvers, M.}
\newblock \bibinfo{title}{A new recombination model for device simulation
  including tunneling}.
\newblock \emph{\bibinfo{journal}{IEEE Transactions on electron devices}}
  \textbf{\bibinfo{volume}{39}}, \bibinfo{pages}{331--338}
  (\bibinfo{year}{1992}).

\bibitem{gonzalez2011analysis}
\bibinfo{author}{Gonzalez, M.~B.} \emph{et~al.}
\newblock \bibinfo{title}{Analysis of the temperature dependence of
  trap-assisted tunneling in \text{Ge pFET} junctions}.
\newblock \emph{\bibinfo{journal}{Journal of The Electrochemical Society}}
  \textbf{\bibinfo{volume}{158}}, \bibinfo{pages}{H955} (\bibinfo{year}{2011}).

\bibitem{sze2021physics}
\bibinfo{author}{Sze, S.~M.}, \bibinfo{author}{Li, Y.} \& \bibinfo{author}{Ng,
  K.~K.}
\newblock \emph{\bibinfo{title}{Physics of semiconductor devices}}
  (\bibinfo{publisher}{John wiley \& sons}, \bibinfo{year}{2021}).

\bibitem{vsvcajev2020temperature}
\bibinfo{author}{{\v{S}}{\v{c}}ajev, P.} \emph{et~al.}
\newblock \bibinfo{title}{Temperature dependent carrier lifetime, diffusion
  coefficient, and diffusion length in \text{Ge$_{0.95}$Sn$_{0.05}$} epilayer}.
\newblock \emph{\bibinfo{journal}{Journal of Applied Physics}}
  \textbf{\bibinfo{volume}{128}}, \bibinfo{pages}{115103}
  (\bibinfo{year}{2020}).

\end{thebibliography}

% \bibliographystylesupp{naturemag}
% \bibliographysupp{methods}

\end{document}